\documentclass{article}
\usepackage{waspaa19,amsmath,graphicx,url,times}
\usepackage{multirow}
\usepackage{color, colortbl, array}
\definecolor{Gray}{gray}{0.9}
\usepackage[noadjust]{cite}

\usepackage{siunitx}

\title{Localization, Detection and Tracking of Multiple Moving Sound Sources with a Convolutional Recurrent Neural Network}

\name{Sharath Adavanne,\thanks{This work has received funding from the European Research Council under the ERC Grant Agreement 637422 EVERYSOUND.}
      Archontis Politis, and
      Tuomas Virtanen}
\address{Audio Research Group, Tampere University, Finland, firstname.lastname@tuni.fi}
\begin{document}

\ninept
\maketitle

\begin{sloppy}

\begin{abstract}
This paper investigates the joint localization, detection, and tracking of sound events using a convolutional recurrent neural network (CRNN). We use a CRNN previously proposed for the localization and detection of stationary sources, and show that the recurrent layers enable the spatial tracking of moving sources when trained with dynamic scenes. The tracking performance of the CRNN is compared with a stand-alone tracking method that combines a multi-source (DOA) estimator and a particle filter. Their respective performance is evaluated in various acoustic conditions such as anechoic and reverberant scenarios, stationary and moving sources at several angular velocities, and with a varying number of overlapping sources. The results show that the CRNN manages to track multiple sources more consistently than the parametric method across acoustic scenarios, but at the cost of higher localization error.
\end{abstract}


\section{Introduction}
\label{sec:intro}
\vspace{-5pt}
Sound event localization, detection, and tracking (SELDT) is the combined task of identifying the temporal onset and offset of potentially temporally-overlapping sound events, recognizing their classes, and tracking their respective spatial trajectory when they are active.
%
%
Performing SELDT successfully provides an automatic description of the acoustic scene that can be employed by machines to interact naturally with their surroundings. Applications such as teleconferencing systems and robots can use this information for tracking the sound event of interest~\cite{Takeda2016_single,Takeda2016_double,Yalta2017,He2018,Butko2011,ghoshal2014}. Furthermore, smart cities and smart homes can use it for audio surveillance~\cite{surveillance_audio, Grobler2017, Wessels2017}.

Localization and detection in static scenes with spatially stationary sources have been studied with different parametric~\cite{Grobler2017, Butko2011, Chakraborty_ICASSP2014, Lopatka2016} and deep neural network (DNN)~\cite{Hirvonen2015} based methods. However, these methods do not employ any temporal modeling required for the tracking of moving sources in dynamic scenes. Recently, we proposed a convolutional recurrent neural network (SELDnet) that was shown to perform significantly better localization and detection than the only other existing DNN-based method~\cite{Hirvonen2015}.  SELDnet's capabilities to localize events in full azimuth and elevation under matched and unmatched acoustic conditions, and without relying on features dependent on specific microphone arrays, were studied and presented in~\cite{Adavanne_JSTSP2018}. However,~\cite{Adavanne_JSTSP2018} studied only static scenes. 

On the other hand, stand-alone tracking methods have been widely studied for both stationary and moving sources based on spatial information only~\cite{Potamitis2004, Valin2007, Roman2008, Zhong2009, Fallon2012, Traa2014a, Schwartz2014},  additional spectral information \cite{Nix2007, Woodruff2013},  or in conjunction with visual information \cite{strobel2001}. Such parametric methods often require manual tuning of multiple parameters corresponding to the scene composition and dynamics, and new sets of parameters have to be identified manually for different sound scenes. Furthermore, tracking usually focuses on distinguishing source trajectories, with no regard to source signal content. In the case of temporally overlapping trajectories, track identities are assigned to individual trajectories, but these identities are not source dependent and are generally re-used for trajectories from different sources across the audio recording. A balance between consistent association and localization determines the tracker's performance in most cases. Alternatively, a detect-before-track approach, as in the proposed SELDnet, circumvents the association problem by first detecting the active sound events, and then assigning a track to each detected event. As long as such a system is able to react to time-varying conditions, with temporally and spatially overlapping sound events from both stationary and moving sources, it is also able to detect and track the sound events of interest.


In this work, we study the multi-source tracking capabilities of a detection and localization system based on our recently proposed SELDnet~\cite{Adavanne_JSTSP2018}. 
We show that training the SELDnet with dynamic scene data results in tracking, in addition to localization and detection. This tracking ability is enabled by the recurrent layers of the SELDnet that can model the evolution of spatial parameters as a sequence prediction task given the sequential features and their spatial trajectory information. We show that the recurrent layers are crucial for tracking, and in comparison to stand-alone trackers they additionally perform detection. Unlike the parametric tracking methods discussed earlier, the recurrent layer is a generic tracking method that learns directly from the data without manual tracker-engineering. Finally, we show that the tracking performance of SELDnet is comparable with stand-alone parametric tracking methods through evaluation on five datasets, representing scenarios with stationary and moving sources at different angular velocities, anechoic and reverberant environments, and different numbers of overlapping sources. The method and all the studied datasets are publicly available\footnote{https://github.com/sharathadavanne/seld-net}.

\section{Method}
\label{sec:methods}
\vspace{-5pt}
An acoustic scene, for example of a \textit{park}, is characterized by the sound events in it such as \textit{dog bark} and \textit{bird call}, produced by their respective sources \textit{dog} and \textit{bird}. In the SELDT task, we aim to acoustically detect the temporal activity of the sound events, and further localize the source producing the sound event in space and track their movement. The detection task produces the onset and offset times for each sound event class when active. Similarly, the tracking produces the spatial trajectory of the sound source movement when the sound event is active.

The SELDnet~\cite{Adavanne_JSTSP2018} is based on a convolutional recurrent neural network architecture, where the direction of arrival (DOA) of sound events is estimated in a regression manner when the sound event is detected to be active. As shown in Figure~\ref{fig:workflow}, the input to SELDnet is a multichannel audio recording, from which a feature extraction block extracts the phase and magnitude components of the spectrogram. The SELDnet maps the spectrogram to two outputs -- sound event detection, and tracking; together they produce the SELDT output. The detection produces temporal activity for sound classes present in the dataset. The tracking produces only one DOA trajectory for each sound class when it is active, i.e., if multiple instances of the same sound class are temporally overlapping, the SELDnet tracks only one or oscillate between the multiple instances. 

\begin{figure}[tp]
  \centering
  \centerline{\includegraphics[height=7cm, width=\linewidth, trim=0.2cm 0.3cm 0.2cm 0.3cm,clip, keepaspectratio]{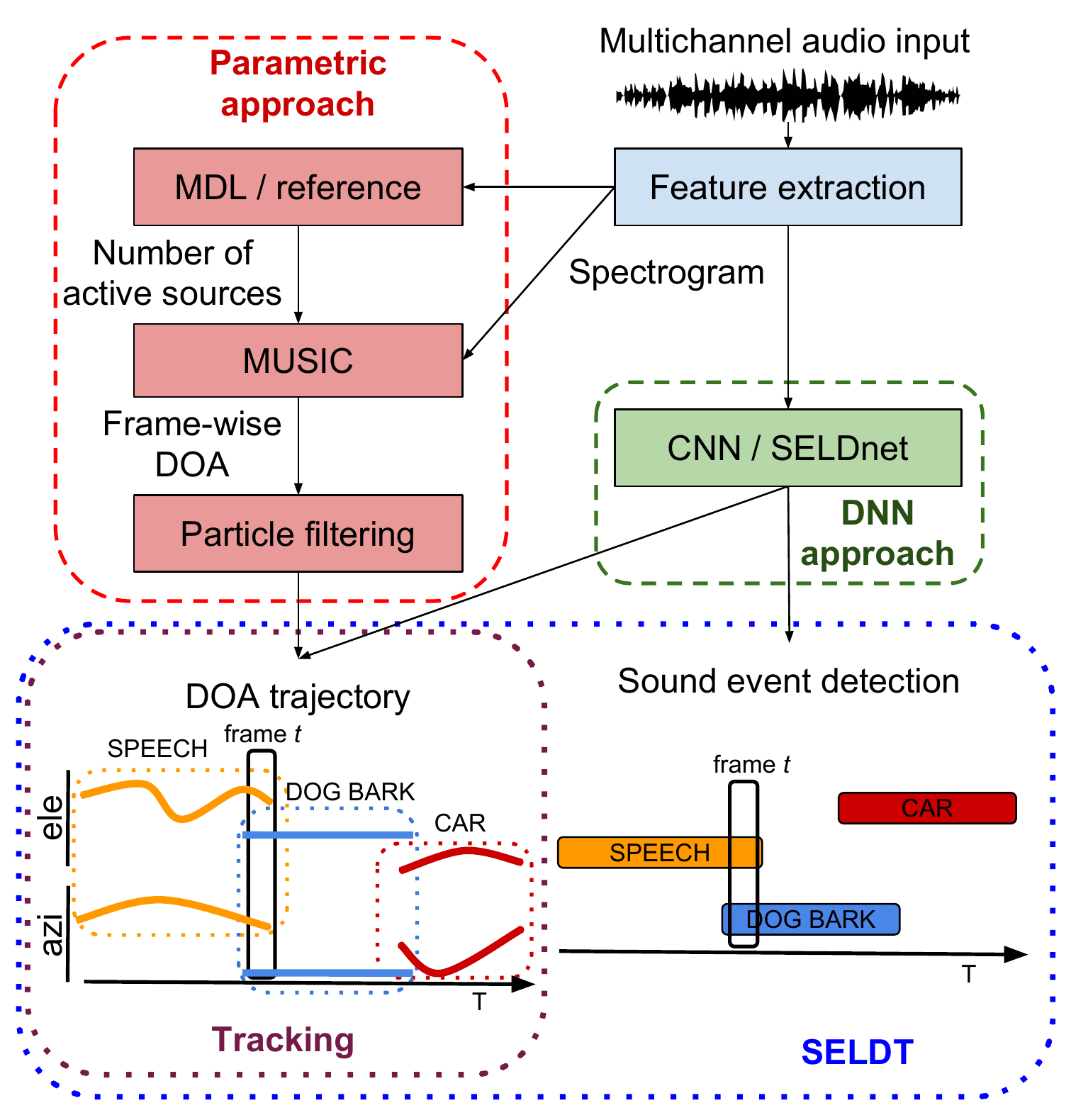}}
  \caption{Workflows for the parametric tracking and DNN-based SELDT approaches. The sound class coloring and naming for the tracking task is only shown here to visualize the concept better. In practice tracking methods do not produce sound class labels as shown in Figure~\ref{fig:MSET}.}
  \label{fig:workflow}
  \vspace{-15pt}
\end{figure}

We use the SELDnet architecture proposed in~\cite{Adavanne_JSTSP2018}, with three convolutional layers of 64 filters each, followed by two-layers of 128-node gated recurrent units. The output of the recurrent layers is fed to two branches of dense layers each with 128 units producing detection and tracking estimates. The convolutional layers in the SELDnet are used as a feature extractor to produce robust features for detection and tracking. Given the spectrogram input of $T$ frames, the convolutional layer produces frame-wise features of similar length $T$. These features are mapped to two frame-wise outputs, detection, and tracking, of the same length $T$, using shared recurrent layers.

The recurrent layers utilize the current input frame along with the information learned from the previous input frames to produce the output for the current frame. This process is similar to a particle filter, which is a popular stand-alone parametric tracker and is also used as a baseline in this paper (see Section~\ref{ssec:baseline}). The particle filter prediction at the current time frame is influenced by both the knowledge accumulated from the previous time frames and the input at the current time frame. For the tracking task of this paper, the particle filter requires the specific knowledge of the sound scene such as the spatial distribution of sound events, their respective velocity ranges when active, and their probability of birth and death. Such concepts are not explicitly modeled in the recurrent layers used in SELDnet, rather they learn equivalent information directly from the input convolutional layer features and corresponding target outputs in the development dataset. In fact, recurrent layers have been shown to work as generic trackers~\cite{gu2017dynamic} that can learn temporal associations of the target source from any sequential input features. Unlike the particle filters that only work with conceptual representations such as frame-wise multiple DOAs for tracking, the recurrent layers work seamlessly with both conceptual and latent representations such as convolutional layer features. 

Finally, by training the recurrent layers in SELDnet using the loss calculated from both detection and tracking, the recurrent layers learn associations between DOAs from neighboring frames corresponding to the same sound class and hence produce the SELDT results. In general, unlike the parametric trackers, the recurrent layers perform similar tracking of the frame-wise DOAs in addition to also detecting their corresponding sound classes. Further, the recurrent layers do not need complicated problem-specific tracker- or feature-engineering that are required by the parametric trackers. A more theoretical relationship between recurrent layers and particle filter is presented in~\cite{choe2017probabilistic}.

\begin{table}[t]
\centering
\caption{Summary of Datasets}
\label{T:datasets}
\resizebox{0.48\textwidth}{!}{%
\begin{tabular}{l|c|c|c}
Sources & Sound scene & Impulse response & Acronym \\ \hline
\multirow{3}{*}{Stationary~\cite{Adavanne_JSTSP2018}} & Anechoic & \multirow{2}{*}{Synthetic} & ANSYN \\ \cline{2-2} \cline{4-4}
 & \multirow{2}{*}{Reverberant} &  & RESYN \\ \cline{3-3} \cline{4-4}
 &  & Real-life & REAL \\ \hline
\multirow{2}{*}{Moving} & Anechoic & Synthetic & MANSYN \\ \cline{2-4}
 & Reverberant & Real-life & MREAL
\end{tabular} 
} \vspace{-15pt}
\end{table}

\section{Evaluation Procedure}
\label{sec:eval}\vspace{-5pt}
\subsection{Datasets}  \vspace{-5pt}

The performance of SELDnet is evaluated on five datasets that are summarized in Table~\ref{T:datasets}. We continue to use the stationary source datasets from our previous work~\cite{Adavanne_JSTSP2018} to evaluate the tracking performance of the parametric tracker that was missing in~\cite{Adavanne_JSTSP2018}, and compare with SELDnet. Further, we create moving-source versions of the ANSYN and REAL datasets to evaluate the performance on moving sources. The studied datasets were synthesized using synthetic and real-life impulse responses for anechoic and reverberant scenarios. The recordings of all datasets are 30 seconds long and captured in the four-channel first-order Ambisonics format \cite{pulkki2017first}. Each dataset has three subsets with no temporally overlapping sources $O1$, maximum two $O2$, and maximum three temporally overlapping sources $O3$. Each of these subsets has three cross-validation splits of 240 recordings for development and 60 for evaluation. All the synthetic impulse response datasets have sound events from 11 classes and DOAs with full azimuth range and elevation range $\in [-60, 60)$. The real-life impulse response datasets have 8 sound event classes and DOAs in full azimuth range and elevation range $\in [-40, 40)$. During synthesis, all the sound events in the stationary source datasets are placed in a spatial grid of \ang{10} resolution for both azimuth and elevation angles. We refer the readers to~\cite{Adavanne_JSTSP2018} for more details on these datasets.  

The anechoic moving source dataset MANSYN has the same sound event classes as ANSYN and is synthesized as follows. Every event is assigned a spatial trajectory on an arc with a constant distance from the microphone (in the range 1-10 m) and moving with a constant angular velocity for its duration. Due to the choice of the ambisonic spatial recording format, the steering vectors for a plane wave source or point source in the far field are frequency-independent. Hence, there is no need for a time-variant convolution or impulse response interpolation scheme as the source is moving; the spatial encoding of the monophonic signal was done sample-by-sample using instantaneous ambisonic encoding vectors for the respective DOA of the moving source. The synthesized trajectories in MANSYN vary in both azimuth and elevation, and are simulated to have a constant angular velocity in the range $\in [\ang{-90}, \ang{90}]/s$ with $\ang{10}/s$ steps. Similarly, the MREAL dataset was synthesized with real-life impulse responses from~\cite{Adavanne_JSTSP2018} that were sampled at \ang{1} resolution along azimuth only. Hence, unlike MANSYN, the sound events in MREAL (that are identical to REAL) have motion only along the azimuth with a constant angular velocity in the range $\in [\ang{-90}, \ang{90}]/s$ and $\ang{10}/s$ steps.

\subsection{Metrics}  \vspace{-5pt}

\label{ssec:metrics}
The evaluation of the SELDT performance is done using individual metrics for detection and tracking identical to~\cite{Adavanne_JSTSP2018}. As the detection metric, we use the F-score and error rate calculated in segments of one-second with no overlap~\cite{metrics}. An ideal detection method will have an F-score of one and an error rate of zero. As the tracking metric, we use two frame-wise metrics: the frame recall and DOA error. The frame recall gives the percentage of frames in which the number of predicted DOAs is equal to the reference. The DOA error is calculated as the angle in degrees between the predicted and reference DOA. In order to associate multiple estimated DOAs with the reference, we use the Hungarian algorithm~\cite{Hungarian} to identify the smallest mean angular distance and use it as DOA error. An ideal tracking method has a frame recall of one and DOA error of zero (see~\cite{Adavanne_JSTSP2018} for more details).

\subsection{Baseline Method}  \vspace{-5pt}

\label{ssec:baseline}

As the baseline method, we use a combination of MUSIC~\cite{Schmidt1986} and an RBMCDA particle filter~\cite{sarkka2007rao} to obtain tracking results in a similar fashion as in~\cite{Valin2007}. The workflow of the baseline method is shown in Figure~\ref{fig:workflow}. MUSIC is a widely used~\cite{Adavanne2018_EUSIPCO,Adavanne_JSTSP2018} subspace-based high-resolution DOA estimation method that can detect multiple narrowband sources. It relies on an eigendecomposition of the narrowband spatial covariance matrix computed from the multichannel spectrogram, and it additionally requires a source number estimate in order to distinguish between a signal and noise subspace. Herein, the number of active sources is taken from the reference of the dataset. To obtain broadband DOA estimates, the narrowband covariance matrices are averaged across three consecutive frames and frequency bins from 50 Hz to 8 kHz. We perform 2D spherical peak-finding on the resulting MUSIC pseudospectrum generated on a 2D angular grid with a \ang{10} resolution for stationary and \ang{1} for moving sources, in both azimuth and elevation. The final output of MUSIC $MUS_{GT}$ is a list of frame-wise DOAs corresponding to the highest peaks equal to the number of active sources in each frame.

The second stage of the parametric method involves a particle filter that produces tracking results by processing the frame-wise DOA information of MUSIC $MUS_{GT}$. The particle filter assumes that the number of sources at each time frame is unknown and tracks them with respect to time using a fixed number of particles. At each time frame, the particle filter receives multiple DOAs and, based on knowledge accumulated from the previous time frames, it assigns each new DOA to one of the existing trajectories, clutter (noise), or a newborn source. Additionally, it also decides if any of the existing trajectories have died. The final output of the particle filter $MUS_{GT}^{PF}$ produces the temporal onset-offset and the DOA trajectory for each of the active sound events. The tracker implementation used in this paper has been publicly released~\footnote{https://github.com/sharathadavanne/multiple-target-tracking}. We refer the reader to~\cite{sarkka2007rao} for the details of this approach. 

\subsection{Experiments}  \vspace{-5pt}

In all our experiments, the baseline particle filter parameters and the sequence length of input spectrogram for SELDnet was tuned using the development set of the respective subset. The performance of the tuned method was tested on the evaluation set of the subset, and the respective metrics averaged across the three cross-validation splits of the subset are reported. 

Unlike the DNN-based method, the parametric method requires additional information on the number of active sources per frame to estimate the corresponding DOAs. However, SELDnet obtains this information from the data itself. In order to have a fair comparison, we used the minimum description length (MDL)~\cite{rissanen1978modeling} principle to estimate the number of sources from the input spectrogram and use it with MUSIC, resulting in the MUSIC output of $MUS_{MDL}$ and the corresponding particle filter output of $MUS_{MDL}^{PF}$. 

Finally, we studied the importance of recurrent layers for the SELDT task by removing them from SELDnet and evaluating the model containing only convolutional and dense layers, referred to as CNN hereafter. The best CNN architecture across datasets had five convolutional layers with 64 filters each. 

\begin{table*}[tp]
\centering
\caption{Evaluation results on different datasets. DE: DOA error, FR: Frame recall, F: F-score, SCOF: Same class overlapping frames}
\label{T:results}
\resizebox{0.9\textwidth}{!}{%
\begin{tabular}{ll|ccc|ccc|ccc|ccc|ccc}
& & \multicolumn{3}{c}{ANSYN} & \multicolumn{3}{c}{RESYN} & \multicolumn{3}{c|}{REAL} & \multicolumn{3}{c}{MANSYN} & \multicolumn{3}{c}{MREAL} \\ \cline{3-17}
\multicolumn{2}{l|}{Tracking results}  & $O1$ & $O2$ & $O3$ & $O1$ & $O2$ & $O3$ & $O1$ & $O2$ & $O3$ & $O1$ & $O2$ & $O3$ & $O1$ & $O2$ & $O3$ \\ \hline
\rowcolor{Gray}
$\mathrm{MUS_{GT}}$ & DE & 1.3 & 5.0 & 12.2 & 21.7 & 28.9 & 32.5 & 15.1 & 33.9 & 44.1 & 0.6 & 14.8 & 28.0  & 16.4 & 34.1 & 43.9 \\ 
\rowcolor{Gray}
$\mathrm{MUS_{GT}^{PF}}$ & DE &0.1 & 1.1 & 2.3 & 4.0 & 5.2 & 6.1 & 3.3 & 8.8 & 12.0 & 0.2 & 4.2 & 8.0 & 3.6 & 8.1 & 11.9 \\
 & FR & 97.0 & 88.5 & 74.3 & 83.8 & 55.6 & 37.3 & 93.0 & 71.0 & 44.7 & 98.7 & 92.3 & 75.1 & 91.0 & 69.9 & 48.3 \\ \rule{0pt}{7pt}\\[-1em]
\multicolumn{17}{l}{Methods estimating the number of active sources directly from input data}\\
\rowcolor{Gray}
$\mathrm{MUS_{MDL}}$ & DE & 0.5 & 14.2 & 24.0 & 22.3 & 31.9 & 38.5 & 25.3 & 36.2 & 44.1 & 4.2 & 17.8 & 28.5 & 26.5 & 35.9 & 44.9 \\
 & FR & 93.9 & \textbf{89.4} & \textbf{86.7} & 61.7 & 45.6 & 52.5 & 53.6 & 35.7 & \textbf{57.5} & 63.8 & 48.1 & 51.85 & 53.4 & 35.2 & \textbf{58.9} \\
\rowcolor{Gray}
$\mathrm{MUS_{MDL}^{PF}}$ & DE & \textbf{0.1} & \textbf{4.4} & \textbf{7.2} & \textbf{6.4} & \textbf{10.6} & \textbf{12.7} & \textbf{9.3} & \textbf{10.9} & \textbf{13.7} &\textbf{ 3.5} & \textbf{6.8} & \textbf{8.0} & \textbf{13.6} & \textbf{11.2} & \textbf{13.6} \\
 & FR & 96.3 & 83.5 & 67.7 & 52.0 & 34.1 & 24.2 & 52.7 & 40.1 & 29.6 & 64 & 49.9 & 39.8 & 58.7 & 34.4 & 27.5 \\  \hline
\rowcolor{Gray}
CNN & DE & 25.7 & 25.2 & 26.9 & 39.1 & 35.1 & 31.4 & 32.0 & 34.9 & 37.1 & 26.1 & 25.8 & 28.2 & 36.6 & 39.3 & 40.2 \\
 & FR & 80.2 & 45.6 & 32.2 & 69.5 & 45.8 & 29.7 & 45.1 & 28.4 & 16.9 & 83.7 & 58.1 & 38.3 & 44.5 & 26.2 & 16.3 \\
\rowcolor{Gray}
SELDnet & DE & 3.4 & 13.8 & 17.3 & 9.2 & 20.2 & 26.0 & 26.6 & 33.7 & 36.1 & 6.0 & 12.3 & 18.6 & 36.5 & 39.6 & 38.5 \\
 & FR & \textbf{99.4} & 85.6 & 70.2 & \textbf{95.8} & \textbf{74.9 }& \textbf{56.4} & \textbf{64.9} & \textbf{41.5} & 24.6 & \textbf{98.5} & \textbf{94.6} & \textbf{80.7} & \textbf{69.6} & \textbf{42.8} & 28.9 \\
\multicolumn{2}{l}{Detection results}  \\ \hline
CNN & ER & 0.52 & 0.46 & 0.51 & 0.44 & 0.45 & 0.54 & 0.52 & 0.51 & 0.51 & 0.59 & 0.47 & 0.48 & 0.46 & 0.49 & 0.52 \\
 & F & 70.1 & 66.5 & 68 & 57 & 54.9 & 42.7 & 50.1 & 49.5 & 48.9 & 65.6 & 62.7 & 60.1 & 55.4 & 50.9 & 48.8 \\
SELDnet & ER & \textbf{0.04} & \textbf{0.16} & \textbf{0.19} & \textbf{0.1} & \textbf{0.29} & \textbf{0.32} & \textbf{0.4} & \textbf{0.49} & \textbf{0.53} & \textbf{0.07} & \textbf{0.1} & \textbf{0.2} & \textbf{0.37} & \textbf{0.45} & \textbf{0.49} \\
 & F & \textbf{97.7} &\textbf{ 89} & \textbf{85.6} & \textbf{92.5} & \textbf{79.6} & \textbf{76.5} & \textbf{60.3} & \textbf{53.1} & \textbf{51.1} & \textbf{95.3} & \textbf{93.2} & \textbf{87.4} & \textbf{64.4} & \textbf{56.4} & \textbf{52.3}\\ \rule{0pt}{7pt}\\[-1em]
 \multicolumn{2}{l|}{SCOF (in \%)} & 0.0 & 4.2 & 12.1 & 0.0 & 4.2 & 12.1 & 0.0 & 7.6 & 23.0 & 0.0 & 3.0 & 9.1 & 0.0 & 7.1 & 20.9
 \end{tabular}%
} \vspace{-15pt}
\end{table*}

\begin{figure}[t]
  \centering
  \centerline{\includegraphics[width=\linewidth, trim=0.4cm 0.35cm 0.35cm 0.25cm,clip]{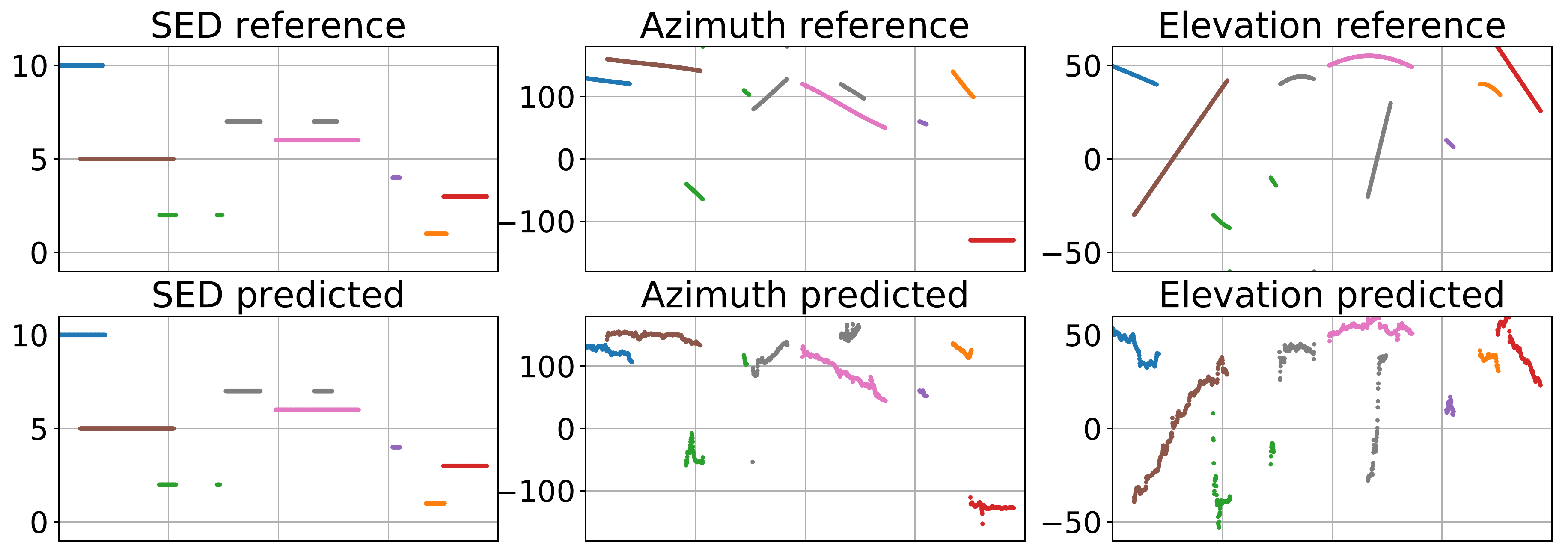}}
    \vspace{-5pt}
  \caption{Visualization of the SELDnet predictions and its respective reference for a MANSYN $O2$ dataset recording. The horizontal-axis of all sub-plots represents the same time frames. The vertical-axis represents sound event class indices for the detection subplots, and DOA azimuth and elevation angles in degrees for remaining subplots. }
  \label{fig:seld}
    \vspace{-15pt}
\end{figure}

\section{Results and discussion}\label{sec:results}
\vspace{-5pt}
On tuning the input sequence length for SELDnet, it was observed that a sequence of 256 frames gave the best scores for the reverberant datasets, and 512 frames gave the best scores for the anechoic datasets. The SELDnet predictions and the corresponding references are visualized in Figure~\ref{fig:seld} for a respective 1000 frame test sequence from MANSYN $O2$ dataset. Each sound class is represented with a unique color across subplots. We see that the detected sound events are accurate in comparison to reference. The DOA predictions are seen to vary around the reference trajectory with a small deviation. This shows that SELDnet can successfully track and recognize multiple overlapping and moving sources.

Figure~\ref{fig:MSET} visualizes the tracking predictions and their respective references for SELDnet and the baseline method $MUS_{GT}^{PF}$. In general, the performance of the two methods is visually comparable. Both the methods are often confused in similar situations, for example in the intervals of 4-5 s, 10-13 s, and 23-25 s.

\begin{figure}[!b]
\vspace{-15pt}
  \centering
  \centerline{\includegraphics[width=\linewidth, trim=0.3cm 0.35cm 0.1cm 0.3cm,clip]{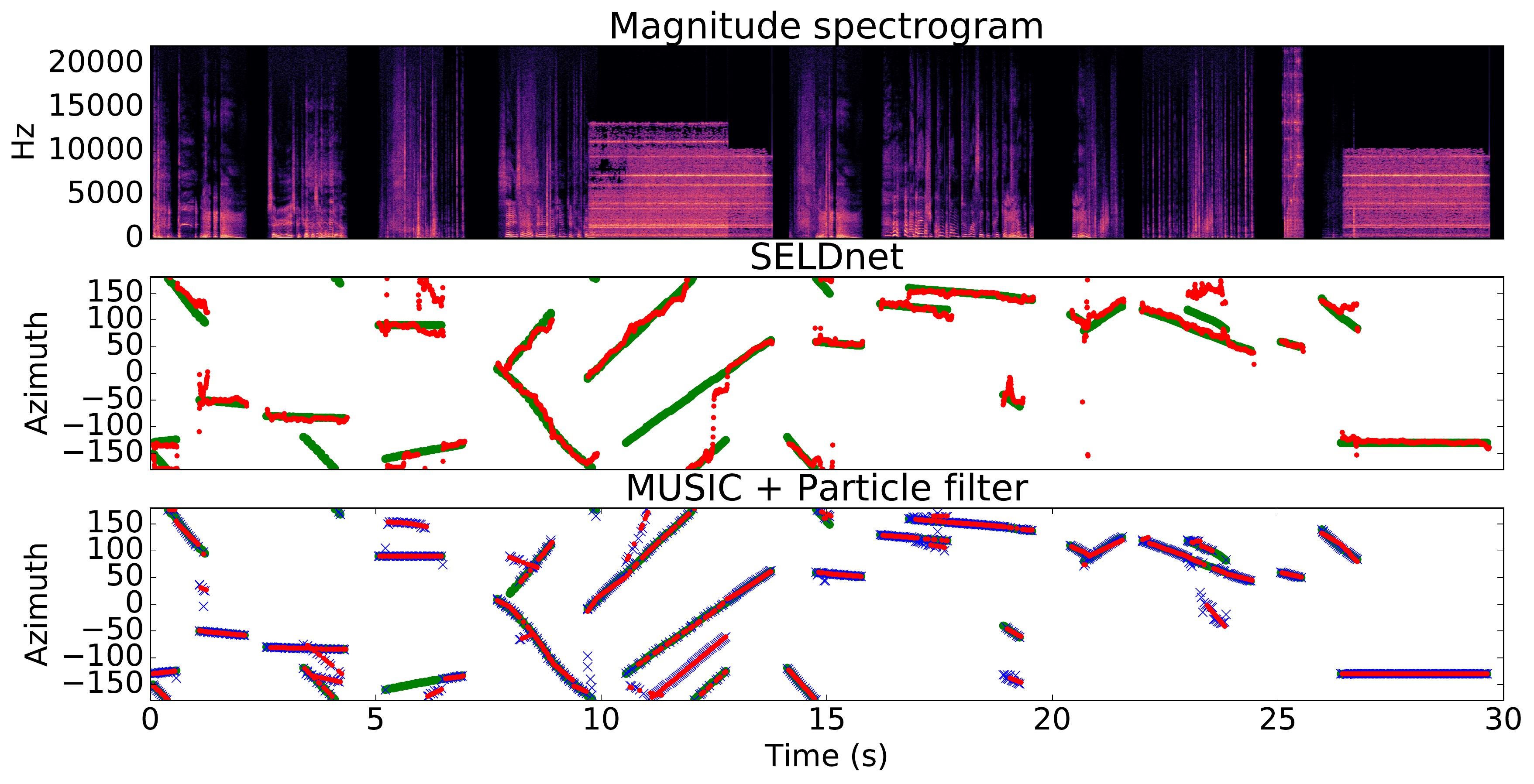}}
  \vspace{-5pt}
  \caption{The tracking results of the two proposed methods are visualized for a MANSYN $O2$ dataset recording. The top figure shows the input spectrogram. The center and bottom figures show the output of SELDnet and $\mathrm{MUS_{GT}^{PF}}$ tracker in red, and the groundtruth in green. The blue crosses in the bottom figure represents the frame-wise DOA output of MUSIC}
  \label{fig:MSET}
\end{figure}

The SELDnet, by design, is restricted to recognize just one DOA for a given sound class. But in real life, there can be multiple instances of the same sound class occurring simultaneously. This is also seen in the datasets studied, the last row (SCOF) in the Table~\ref{T:results} presents the percentage of frames in which the same class is overlapping with itself. In comparison, the parametric method has no such restriction by design and can potentially perform better than SELDnet in these frames (even though, highly correlated sound events coming from different DOAs can easily degrade the performance of parametric methods such as MUSIC). The performance of the two methods in such a scenario can be observed in the 10-13 s interval of Figure~\ref{fig:MSET}. The SELDnet tracks only one of the two sources, while the parametric method tracks both overlapping sources and introduces an additional false track between the two trajectories.

Table~\ref{T:results} presents the quantitative results of the studied methods. The general trend is as follows. The higher the number of overlapping sources, the lower the tracking performance by both SELDnet and the parametric method§. Across datasets, the DOA error improves considerably with the use of the temporal particle filter tracker, but at the cost of lower frame recall. By using MDL instead of reference information for the source number, the overall performance of the parametric approach reduces ($MUS_{GT}^{PF} > MUS_{MDL}^{PF}$). This reduction is especially observed in the frame recall metric, that drops significantly for reverberant and moving source scenario datasets, indicating the need for more robust source detection and counting schemes. 

The frame recall of SELDnet is observed to be consistently better than $MUS_{MDL}^{PF}$, but the DOA estimation is poorer across datasets. A similar relationship is observed between SELDnet and $MUS_{GT}^{PF}$ for all the datasets generated with simulated impulse responses, while for the real-life impulse response datasets the frame recall of SELDnet is poorer than $MUS_{GT}^{PF}$. That could indicate the need for more extensive learning for real-life impulse response datasets, with larger datasets and stronger models.

Using recurrent layers definitely helps the SELDT task. It was observed from visualizations that the tracking performance by the CNN was poor, with spurious and high variance DOA tracks, thus resulting in poor DOA error across datasets as seen in Table~\ref{T:results}. This suggests that the recurrent layers are crucial for SELDT task and perform a similar task as an RBMCDA particle filter of identifying the relevant frame-wise DOAs and associating these DOAs corresponding to the same sound class across time frames.

\section{Conclusion}\vspace{-5pt}
In this paper, we presented the first deep neural network based method, SELDnet, for the combined tasks of detecting the temporal onset and offset time for each sound event in a dynamic acoustic scene, localizing them in space and tracking their position when active, and finally recognizing the sound event class. The SELDnet performance was evaluated on five different datasets containing stationary and moving sources, anechoic and reverberant scenarios, and a different number of overlapping sources. It was shown that the recurrent layers employed by the SELDnet were crucial for the tracking performance. Further, the tracking performance of SELDnet was compared against a stand-alone parametric method based on multiple signal classification and particle filter. In general, the SELDnet tracking performance was comparable to the parametric method and achieved a higher frame recall for tracking but at a higher angular error. 
\bibliographystyle{IEEEtran}
\bibliography{refs19}

\end{sloppy}
\end{document}